\documentclass[copyright]{eptcs}
\providecommand{\event}{TARK 2015}
\usepackage{breakurl}
\usepackage{pifont}
\usepackage{graphicx}
\usepackage{latexsym}
\usepackage{amsfonts}
\usepackage{amsmath}
\usepackage{amssymb}

\usepackage{float}
\newcommand{\p}{{\rm P}}

\newtheorem{theorem}{Theorem}
\newtheorem{corollary}[theorem]{Corollary}
\newcommand\qedblob{\ding{113}}
\def\literalqed{{\ \nolinebreak\hfill\mbox{\qedblob\quad}}}
  \newtheorem{lemma}[theorem]{Lemma}
  \newtheorem{observation}[theorem]{Observation}
  \newtheorem{definition}[theorem]{Definition}
  \newtheorem{example}[theorem]{Example}
  \newtheorem{construction}[theorem]{Construction}
\showhyphens{}

\newcommand{\thetatwo}{\ensuremath{\Theta_2^p}}
\newenvironment{proofs}{\noindent{\bf Proof.}\hspace*{1em}}{\literalqed\bigskip}
\def\extsp{$\exists$-single-peaked} %
\def\itextsp{$\exists$-single-peaked} %

\title{Single-Peaked Consistency for Weak Orders Is Easy}
\author{
Zack Fitzsimmons
\institute{College of Computing and Information Sciences\\
Rochester Institute of Technology\\
Rochester, NY 14623, USA}
\email{zmf6921@rit.edu}
}
\def\titlerunning{Single-Peaked Consistency for Weak Orders Is Easy}
\def\authorrunning{Z. Fitzsimmons}

\begin{document}
\sloppy

\maketitle
\begin{abstract}
In economics and social choice single-peakedness is one of the most
important and
commonly studied models for preferences.
It is well known that
single-peaked consistency for total orders is in \p.
However in practice
a preference profile is not always comprised of total orders.
Often voters have
indifference between
some of the candidates.
In a weak preference order indifference must be transitive.
We show that
single-peaked consistency for weak orders is in \p\
for three different variants of
single-peakedness for weak orders.
Specifically, we consider Black's original definition of single-peakedness for
weak orders,
Black's definition of single-plateaued preferences,
and the existential model recently introduced by Lackner.
We accomplish our results by transforming each of these single-peaked 
consistency
problems to the problem of determining if a 0-1 matrix has the
consecutive ones property.
\end{abstract}

\section{Introduction}

Single-peakedness is one of the most important and commonly examined
domain restrictions on
preferences in economics and social choice.
The study of single-peaked preferences in computational social
choice is often restricted to total orders,
but in practical settings
voters often have some degree of indifference %
in their
preferences. This is seen in the online repository
\textsc{PrefLib}, which contains several %
datasets comprised %
of voters with various degrees of partial %
preferences, many of
which are
weak orders~\cite{mat-wal:b:preflib}.
Additionally, some
election systems are defined for weak orders, e.g.,
the Kemeny rule~\cite{kem:j:no-numbers} and
the Schulze rule~\cite{sch:j:clone-independent-new},
or can be easily extended for weak orders.

Single-peaked preferences were introduced by
Black~\cite{bla:j:rationale-of-group-decision-making} and
they model the preferences of a collection of voters
with respect to a one-dimensional
axis, i.e., a total ordering of the candidates.
Each voter in a single-peaked electorate has a single most preferred
candidate (peak) on the axis and the farther that a candidate is from the voter's
peak the less preferred they are by the voter.
Black extended his model to
single-plateaued preferences, which models the preferences of a collection of
voters in a similar way, but allows voters to have multiple most
preferred candidates (an indifference plateau) in their preferences~\cite[Chapter 5]{bla:b:polsci:committees-elections}.
We mention that the definition of single-peaked preferences from
Fishburn~\cite[Chapter 9]{fis:b:theory} for weak orders is the same as Black's
definition of single-plateaued preferences.
Elections where the voters have single-peaked preferences
over the candidates
have many desirable properties in economics and social choice, e.g.,
the majority relation is transitive~\cite{bla:j:rationale-of-group-decision-making}
and there exist strategy-proof voting rules~\cite{mou:j:strategy-proof}.
Additionally, computational problems often become easier when preferences
are single-peaked.
For example, 
when voters %
in an election
have single-peaked
(or even \emph{nearly} single-peaked)
preferences the complexity of determining if
a manipulative action exists often becomes
easy~\cite{fal-hem-hem-rot:j:single-peaked-preferences,fal-hem-hem:j:nearly-sp} and
determining the winner
for Dodgson and Kemeny elections becomes easy~\cite{bra-bri-hem-hem:c:sp2}
when it is $\thetatwo$-complete in general~\cite{hem-hem-rot:j:dodgson,hem-spa-vog:j:kemeny}.

The problem of single-peaked consistency is to determine if an axis
exists such that the preferences of a collection of voters are single-peaked.
The first paper to computationally study single-peaked consistency
for partial preferences was Lackner~\cite{lac:c:incomplete-sp-aaai},
where a partial order is said to
be single-peaked with respect to an axis if it can be extended to
a total order that is single-peaked with respect to that axis. %
For clarity we refer to this as existentially single-peaked, or
\extsp , throughout this paper.
Lackner presents algorithms and complexity results
for determining the \extsp\
consistency for
preference profiles of varying degrees
of partial %
preferences, including top orders, weak orders,
local weak orders, and
partial orders.
Lackner shows that
if a given preference profile contains
an implicitly specified total order (which is not guaranteed to exist)
then \extsp\ consistency for weak orders
is in
\p~\cite{lac:c:incomplete-sp-aaai}. %
Lackner also shows that the
general
case of \extsp\ consistency for top orders (weak orders with all
indifference between last-ranked candidates) is in \p~\cite{lac:c:incomplete-sp-aaai}.
The complexity of
the general case of \extsp\ consistency
for weak orders was explicitly left as the main open problem %
in Lackner~\cite{lac:c:incomplete-sp-aaai} and we show in this paper that it is
in \p.
We show that an algorithm to determine
if a 0-1 matrix has the
consecutive ones property
can be used to determine the single-peaked, %
single-plateaued, and %
\extsp\ consistency for %
weak orders \emph{without} requiring an implicitly specified
total order.
So given a preference profile of weak orders not only can we determine
if it is single-peaked, single-plateaued, or \extsp , we can
find all consistent axes by using %
the PQ-tree 
algorithm %
for determining if a 0-1 matrix has the
consecutive ones property~\cite{boo-lue:j:pqtrees}. %
This algorithm
was previously %
used to determine 
the single-peaked consistency for total orders %
by
Bartholdi and
Trick~\cite{bar-tri:j:stable-matching-from-psychological-model}
and to determine the single-crossing consistency for total orders
by Bredereck et al.~\cite{bre-che-woe:j:char-single-crossing}.
The model of single-crossing preferences
is another domain restriction~\cite{mir:j:single-crossing} and
its corresponding consistency problem for total orders was first shown to be
in \p\ by Elkind et al.~\cite{elk-fal-sli:c:decloning}.
We also mention that after single-peaked consistency for total orders
was shown to be in \p, both
Escoffier et al.~\cite{esc-lan-ozt:c:single-peaked-consistency} 
and Doignon and Falmagne~\cite{doi-fal:j:unidimensional-unfolding}
independently found faster direct algorithms. %

This paper is organized as follows.
In Section~\ref{sec:preliminaries} we define the types of partial preferences
studied, the
different variants %
of single-peakedness, and the consecutive ones 
matrix problem.
We present our results in Section~\ref{sec:results}, %
which is split into three
sections with each
corresponding to a different variant of %
single-peakedness. Section~\ref{sec:extsp} contains our results
for \extsp\ preferences, Section~\ref{sec:splat} for
single-plateaued preferences, and Section~\ref{sec:sp} for
single-peaked preferences.
In each of these sections
we redefine the %
variant %
of single-peakedness using forbidden substructures and describe
the transformation
from its consistency problem to the problem of determining if
a 0-1 matrix has the consecutive ones property. %
We
conclude in Section~\ref{sec:conclusion} by summarizing our results
and
stating some possible directions for future work. %

\section{Preliminaries}
\label{sec:preliminaries}

A \emph{preference order}, $v$, is an ordering of the elements of a finite
candidate set, $C$. %
A multiset of preference orders, $V$, is called a
\emph{preference profile}.
(We sometimes refer to each $v$ as a voter with a corresponding
preference order.)
A \emph{partial order} is
a transitive and reflexive binary relation on a set.
A \emph{weak order} is a partial order that is complete. %
a \emph{top order} is a weak order where all indifference is between candidates
ranked last, and %
a \emph{total order} is a weak order with no indifference between
candidates. 

\begin{example}\normalfont
Given the set of candidates %
$\{a,b,c,d\}$, an example of a total order is $(a > b > d > c)$, 
an example of a weak order is $(a \sim c > d > b)$,
and an example of a top order is $(a > c > b \sim d)$,
where ``$\sim$'' is used to denote indifference %
between candidates.
\end{example}

We focus on weak orders since they model natural
cases where voters are not able to discern between
two candidates or where they view them as truly equal. 
Allowing each voter to
state a weak preference order still requires that they specify each candidate
in their order, but gives them the ability to have multiple
candidates at each position.

It is very natural for election systems to be defined for weak orders.
The Kemeny rule and Schulze rule are defined for
weak orders~\cite{kem:j:no-numbers,sch:j:clone-independent-new},
and clearly election systems based on
pairwise comparisons (e.g., Copeland) can be used to evaluate a preference
profile of partial votes. %
The Borda
count can be extended for top orders~\cite{eme:j:partial-borda}
and a
recent paper has even explored the complexity of the manipulation problem for
such extensions to the Borda count and defined additional
extensions for election systems to be defined for top
orders~\cite{nar-wal:c:partial-vote-manipulation}.

\subsection{Variants of Single-Peakedness}\label{sec:variants}

In our definitions of each variant of single-peakedness we refer to a total
ordering of the set of candidates that each preference profile is consistent
with as an axis $A$.
Like Bartholdi and Trick~\cite{bar-tri:j:stable-matching-from-psychological-model},
who were the first to show single-peaked consistency for total orders in \p,
we say that a preference order $v$ is strictly increasing (decreasing) along
a segment $X$ of $A$ if each candidate in $X$ is strictly preferred to each
candidate to its left (right) in $X$.
Similarly, we say that a preference order is increasing (decreasing) along
a segment $X$ of $A$ if each candidate in $X$ is strictly preferred or
ranked indifferent to each candidate to its left (right) in $X$.
When we say
that a preference order $v$ is remaining constant along a segment, then all
candidates in that segment are ranked indifferent to each other. %

We begin our discussion of single-peaked preferences
by stating the
definition of single-peaked preferences for total orders. %
We use the definition found in the work by Bartholdi and
Trick~\cite{bar-tri:j:stable-matching-from-psychological-model}.
\begin{definition}\label{def:blacksp-total}
A preference profile $V$ of total orders is \emph{single-peaked} with respect to
an axis $A$ if for every $v \in V$, %
$A$ can be split
at the most preferred candidate (peak) of $v$ into two segments $X$ and $Y$
(one of which can be empty) such that
$v$ has strictly increasing
preferences along $X$ and $v$ has strictly
decreasing preferences along $Y$.
\end{definition}

We now define each of the three variants of single-peaked preferences
for weak orders that we study in this paper and present an
example of each in Figure~\ref{fig:exp}.

\subsubsection{Single-Peaked Preferences}

Single-peaked preferences for weak orders can be defined in the %
same way as single-peaked preferences for total orders.
\begin{definition}\label{def:blacksp}
A preference profile $V$ of weak orders is \emph{single-peaked} with respect to
an axis $A$ if for every $v \in V$, %
$A$ can be split
at the most preferred candidate (peak) of $v$ into two segments
$X$ and $Y$ (one of which can be empty) such that
$v$ has strictly increasing preferences along $X$ and $v$ has
strictly decreasing preferences along $Y$.
\end{definition}

Notice that for a weak preference
order to be single-peaked it must have a single most preferred candidate and can
only contain indifference between at most two candidates at each
position. Otherwise the segments $X$ and $Y$ referred to in
Definition~\ref{def:blacksp} would not be \emph{strictly}
increasing/decreasing.
We define the corresponding problem of single-peaked consistency for weak
orders below.

\begin{description}
\item[Given:] A preference profile $V$ of weak orders and a set of candidates $C$.

\item[Question:] Does there exist an axis $A$ such that $V$ is single-peaked
  with respect to $A$?
\end{description}

\subsubsection{Single-Plateaued Preferences}

A slightly weaker restriction than single-peakedness for weak %
orders %
is
single-plateauedness~\cite[Chapter 5]{bla:b:polsci:committees-elections}.
Single-peaked and single-plateaued preferences are closely related domain
restrictions and
Barber{\`a}~\cite{bar:j:indifference-domain} discusses how the
amounts of indifference permitted in these restrictions impact their
properties.

Building upon the %
definition for single-peaked preferences, we state
a definition for single-plateaued preferences.

\begin{definition}\label{def:splat}
A preference profile $V$ of weak orders is \emph{single-plateaued}
with respect to
an axis $A$ if for every $v \in V$, %
$A$ can 
be split into three  segments $X$, $Y$, and $Z$ 
($X$ and $Z$ can each be empty)
where $v$'s most preferred candidates are $Y$, %
$v$ has strictly increasing preferences along $X$,
and $v$ has strictly decreasing preferences along $Z$.
\end{definition}
 
We define the corresponding problem of single-plateaued
consistency for weak orders below.
 
\begin{description}
\item[Given:] A preference profile $V$ of weak orders and a set of candidates $C$.

\item[Question:] Does there exist an axis $A$ such that $V$ is single-plateaued
  with respect to $A$?
\end{description}

\subsubsection{Existentially Single-Peaked Preferences}

So far we have considered the given preference orders as the true
preferences of the voters.
One approach to dealing with
partial preferences is to assume that
voters have an underlying total preference
order and consider extensions
of their preferences to total orders (see,
e.g.,~\cite{kon-lan:c:incomplete-prefs}).
This is
the approach taken by Lackner for the existential model of
single-peakedness~\cite{lac:c:incomplete-sp-aaai}.

\begin{definition}\label{def:ext-sp}\cite{lac:c:incomplete-sp-aaai}
    A preference profile
    $V$ of weak orders is \emph{\extsp}\ with respect to an axis $A$
    if for every
    $v \in V$, $v$ can be extended to a total order $v'$
     such that the profile
    $V'$ of total orders is single-peaked with respect to $A$.
\end{definition}

We can restate Definition~\ref{def:ext-sp} %
without referring to extensions to better see how it relates to
single-peaked and single-plateaued preferences. %

\begin{observation}\label{def:ext-alt}
A preference profile $V$ of weak orders is \emph{\extsp}\ with respect to
an axis $A$ if and only if for every $v \in V$, %
$A$ can 
be split into three segments $X$, $Y$, and $Z$  ($X$ and $Z$ can each be empty)
where $v$'s most preferred candidates are $Y$, %
$v$ has increasing preferences along $X$,
and $v$ has decreasing preferences along $Z$.
\end{observation}
 
We define the corresponding problem of \extsp\ consistency for weak
 orders below.
 
\begin{description}
\item[Given:] A preference profile $V$ of weak orders and a set of candidates $C$.

\item[Question:] Does there exist an axis $A$ such that $V$ is \extsp\
  with respect to $A$?
\end{description}

\begin{figure}[t]
\centering
\includegraphics{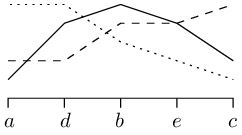}
\caption{The solid line represents the single-peaked preference order
$(b > d \sim e > c >a)$, the dotted line represents the single-plateaued 
preference order $(a \sim d > b > e > c)$, and the dashed line 
represents the \extsp\ preference order $(c > b \sim e > d \sim a)$.}
\label{fig:exp}
\end{figure}

Figure~\ref{fig:exp} illustrates an example of each variant of
single-peakedness for weak orders described
where each preference order is consistent with respect to the axis $A = a < d < b < e < c$.
In Figure~\ref{fig:exp}
the preference order
$(b > d \sim e > c > a)$ is single-peaked,
single-plateaued, and \extsp. %
The preference order $(a \sim d > b > e > c)$ is single-plateaued
and \extsp , but not %
single-peaked since it has more than one most preferred candidate.
The preference order $(c > b \sim e > d \sim a)$ is \extsp\
and not single-plateaued or single-peaked since
it is not strictly increasing to its most preferred candidate(s).

We conclude our discussion of these variants of single-peakedness for weak
orders by stating several observations.

First we show that there exists
an \extsp\ consistent preference profile that does not have
a transitive majority relation. We say that a majority relation is transitive if
when $x > y$ and $y > z$ by majority, then $x > z$ by majority.
Note that 
single-peaked and single-plateaued preferences both have transitive majority
relations~\cite{bla:j:rationale-of-group-decision-making,bla:b:polsci:committees-elections}. %
Consider the preference profile $V$ comprised of the following
five voters from Table~9.1 in Fishburn~\cite{fis:b:theory}.
\[\begin{matrix}
  v_1   & (b > a > c)\\
  v_2,v_3 & (c > b > a)\\
  v_4,v_5 & (a > b \sim c)\\
\end{matrix}\]
When we evaluate this preference profile under the
simple majority rule %
where $x > y$ by simple majority if more voters state $x > y$ than
$y > x$,
then $V$ has the %
majority
cycle $a > c > b > a$~\cite{fis:b:theory}.
Clearly $V$ is \extsp\ consistent with respect to the axis $A = a < b < c$, so
we can make the following observation.

\begin{observation}
There exists a preference profile of weak orders that is \extsp\
and does not have a transitive majority relation.
\end{observation}

The existential model for single-peakedness %
considers
the existence of a single extension of %
the preferences of all of the voters to total orders.
We briefly consider %
the case %
where all extensions to total orders
must be single-peaked and make two observations.

\begin{observation}
If a preference profile of weak orders is single-peaked then all
extensions of the preferences %
to total orders
are also single-peaked.
\end{observation}

\begin{observation}
If a preference profile of weak orders is single-plateaued
and  each preference order has at most two most preferred candidates,
then all extensions of the preferences %
to total orders
are single-peaked.
\end{observation}

\subsection{Consecutive-Ones Matrices}

All of
our polynomial-time results are due to %
transformations %
to the following problem of determining if a
0-1 matrix has the consecutive ones property. %

\begin{description}
 \item[Given:] A 0-1 matrix $M$.
  \item[Question:] Does there exist a permutation of the columns of $M$ such
      that in each row all of the 1's are consecutive?
\end{description}

The above problem
was shown to be in \p\
by Fulkerson and Gross~\cite{ful-gro:j:incidence-matrices}.
Booth and Lueker~\cite{boo-lue:j:pqtrees}
improved on this result by finding a linear-time algorithm through the
development and use of the novel PQ-tree
data structure,
which contains all possible permutations of the columns of a matrix
such that all of the 1's are consecutive in each row.

\section{Results}\label{sec:results}

The following three sections consist of our results and they are
structured as
follows. We examine each variant %
of single-peakedness starting with the weakest
restriction and ending
with the strongest.
When we examine each restriction we present an alternative
definition of the variant %
of single-peakedness using forbidden substructures and the transformation
to the problem of determining if a 0-1 matrix has the consecutive ones property.

\subsection{Existentially Single-Peaked Consistency}\label{sec:extsp}

The most general of the three variants mentioned in
Section~\ref{sec:variants} is the model of \extsp\ preferences. 
The construction and corresponding
proof will be the basis for showing that single-peaked and
single-plateaued consistency for weak orders are each also in \p.

Given an axis $A$ and a preference order $v$, if $v$ is
\extsp\ with respect to $A$ then $v$ cannot have strictly decreasing
and then strictly increasing preferences with respect to $A$.
Following the terminology used by
Lackner~\cite{lac:c:incomplete-sp-aaai}, we refer to this as a
v-valley.

\begin{definition}\label{def:v-valley}
   A preference order $v$ over a candidate set $C$ contains a
   \emph{v-valley} with respect to an axis $A$ if there exist
   candidates %
   $a,b,c \in C$ such that $a < b < c$ in $A$ and $(a > b)$ and $(c > b)$
   in $v$.
\end{definition}

Using the v-valley substructure we can state the following lemma, which
will simplify our argument used in the proof of Theorem~\ref{thm:main}.

\begin{lemma}\label{lem:extv}\cite{lac:c:incomplete-sp-aaai}
Let $V$ be a preference profile of weak orders. $V$ is \extsp\ with
respect to an axis $A$ if and only if no preference order $v \in V$ contains 
a v-valley with respect to $A$.
\end{lemma}

To construct a matrix from a preference profile of weak orders, we apply
essentially the same transformation
as used Bartholdi and
Trick~\cite{bar-tri:j:stable-matching-from-psychological-model}
for total orders (see Example~\ref{ex:extsp}).
We describe the construction below. %

\begin{construction}\label{con:mat} %
Let $V$ be a preference profile of weak orders over candidate set $C$. %
For each $v \in V$ construct a ($\|C\|-1) \times \|C\|$ matrix $X_v$.
Each column of $X_v$ corresponds to a candidate in $C$. For each candidate
$c \in C$ let $k$ be the number of candidates that are strictly
preferred to $c$ in $v$ and %
let the corresponding column
in matrix $X_v$ contain $k$ 0's starting at row one, with the 
remaining
entries filled with 1's.
All $\|V\|$ of the matrices are %
row-wise concatenated to yield the
$(\|V\| \cdot (\|C\|-1)) \times \|C\|$ matrix~$X$.
\end{construction}
The main difference in our construction is
that we have one fewer row in each of the individual preference matrices.
In the construction used by Bartholdi and
Trick~\cite{bar-tri:j:stable-matching-from-psychological-model},
given a preference order $v$ over a set of candidates $C$,
for all $a,b \in C$, $(a > b)$ in $v$ if and only if
the number of 1's in the column corresponding to
$a$ is greater than the number of 1's in the column
corresponding to $b$ in $v$'s corresponding
individual preference
matrix. %
Notice that this still holds for our construction.
The polynomial-time results for \extsp\ consistency for weak orders and local
weak orders proved in Lackner~\cite{lac:c:incomplete-sp-aaai}
require that the given preference profile contains a \emph{guiding order}, i.e., an
implicitly specified total order. %
Given a preference profile $V$,
a guiding order can be constructed iteratively in the following way.
If there exists a $v \in V$ such that the last ranked candidate in %
$v$ is not ranked indifferently %
with any other candidate, then
that candidate is appended to the top of the guiding order.
This is then repeated on the preference profile restricted to the
candidates not yet added
to the guiding order until either the guiding order is a total order or
there is no $v \in V$
with a unique %
last ranked candidate,
the case where no guiding order exists~\cite{lac:c:incomplete-sp-aaai}.
Observe that
if a given preference profile is \extsp\ then it remains \extsp\ if a
guiding order is added %
as an additional preference order~\cite{lac:c:incomplete-sp-aaai}.
It is important to point out that our results do not
depend on the existence of a guiding order in a preference profile.
Below we show how Construction~\ref{con:mat} is applied to a preference profile
of weak orders that is \extsp.

\begin{example}\label{ex:extsp}\normalfont
Consider the preference profile $V$ that consists of the
preference orders $v$ and $w$.
Let the preference order $v$ be $(a \sim c > b > e \sim d > f)$ and
the preference order
$w$ be $(a > b > c > e \sim d > f)$.
Notice that %
$V$ does not contain a guiding order,
which is required by the polynomial-time algorithm for weak orders
found in Lackner~\cite{lac:c:incomplete-sp-aaai}.
\[X_v =    \stackrel{\begin{matrix} a & b & c & d & e & f\\ \end{matrix}}{
   \begin{bmatrix}
   1 & 0 & 1 & 0 & 0 & 0\\
   1 & 0 & 1 & 0 & 0 & 0\\
   1 & 1 & 1 & 0 & 0 & 0\\
   1 & 1 & 1 & 1 & 1 & 0\\
   1 & 1 & 1 & 1 & 1 & 0\\
    \end{bmatrix}}
\hspace{0.25\textwidth}
X_w =    \stackrel{\begin{matrix} a & b & c & d & e & f\\ \end{matrix}}{
   \begin{bmatrix} 
   1 & 0 & 0 & 0 & 0 & 0\\
   1 & 1 & 0 & 0 & 0 & 0\\
   1 & 1 & 1 & 0 & 0 & 0\\
   1 & 1 & 1 & 1 & 1 & 0\\
   1 & 1 & 1 & 1 & 1 & 0\\
\end{bmatrix}}\]
We then row-wise concatenate $X_v$ and $X_w$ to construct $X$.
\[ X = 
   \stackrel{\begin{matrix} a & b & c & d & e & f\\ \end{matrix}}{
   \begin{bmatrix}
   1 & 0 & 1 & 0 & 0 & 0\\
   1 & 0 & 1 & 0 & 0 & 0\\
   1 & 1 & 1 & 0 & 0 & 0\\
   1 & 1 & 1 & 1 & 1 & 0\\
   1 & 1 & 1 & 1 & 1 & 0\\
   1 & 0 & 0 & 0 & 0 & 0\\
   1 & 1 & 0 & 0 & 0 & 0\\
   1 & 1 & 1 & 0 & 0 & 0\\
   1 & 1 & 1 & 1 & 1 & 0\\
   1 & 1 & 1 & 1 & 1 & 0\\
    \end{bmatrix}}
\hspace{0.25\textwidth}
 X' = 
   \stackrel{\begin{matrix} b & a & c & d & e & f\\ \end{matrix}}{
   \begin{bmatrix}
    0 & 1 & 1 & 0 & 0 & 0\\
    0 & 1 & 1 & 0 & 0 & 0\\
    1 & 1 & 1 & 0 & 0 & 0\\
    1 & 1 & 1 & 1 & 1 & 0\\
    1 & 1 & 1 & 1 & 1 & 0\\
    0 & 1 & 0 & 0 & 0 & 0\\
    1 & 1 & 0 & 0 & 0 & 0\\
    1 & 1 & 1 & 0 & 0 & 0\\
    1 & 1 & 1 & 1 & 1 & 0\\
    1 & 1 & 1 & 1 & 1 & 0\\
    \end{bmatrix}}\]
Next, we permute the columns of $X$ so that in each row all of the 1's
are consecutive to yield $X'$.
Observe that $V$ is \extsp\ with respect to $b < a < c < d < e < f$,
the ordering of the
columns of $X'$ as its axis.
\end{example}

We now show that \extsp\ consistency for weak orders and
the problem of determining if the constructed 0-1 matrix has the consecutive
ones property %
are equivalent using Lemma~\ref{lem:extv} and Construction~\ref{con:mat}. %

\begin{theorem}\label{thm:main}
A preference profile $V$ of weak orders is \itextsp\ consistent if and
only if
the matrix $X$,
constructed using Construction~\ref{con:mat},
has the consecutive ones property. %
\end{theorem}

\begin{proofs}
Let $V$ be a preference profile of weak orders.
Essentially the same argument as used by Bartholdi and
Trick~\cite{bar-tri:j:stable-matching-from-psychological-model} holds.

If $V$ is \extsp\ with respect to an axis $A$
then by Lemma~\ref{lem:extv} we know that
no preference order $v \in V$ contains a v-valley
with respect to $A$.
When the columns of the matrix $X$ are
permuted to correspond to the axis $A$
no row will  contain the sequence
$\cdots 1 \cdots 0 \cdots 1 \cdots$
since this corresponds to a preference order that strictly decreases
and then strictly increases along the axis $A$ (a v-valley).
Therefore $X$ has the consecutive ones property. %

For the other direction suppose that
$V$ is not \extsp, then by Lemma~\ref{lem:extv} we know that for every
possible axis there exists a preference order $v \in V$ such that $v$ contains a
v-valley with respect to that axis.
So every permutation of the columns of $X$ will
correspond to an axis where some preference order has a v-valley.
As stated in the other direction, a v-valley
corresponds to a row containing the sequence
$\cdots 1 \cdots 0 \cdots 1 \cdots$ so clearly $X$ does not have the
consecutive ones property.

The only difference from the argument used by Bartholdi and
Trick~\cite{bar-tri:j:stable-matching-from-psychological-model} for
total orders is
that in our case the preference orders can remain constant at the peak and
at points on either side of the peak. The same argument still applies since
by Lemma~\ref{lem:extv} the absence of v-valleys with respect to
an axis is equivalent to a profile of weak orders being \extsp\ with
respect to that axis.~\end{proofs}

\begin{corollary}\label{thm:extsp-consist}
$\exists$-Single-peaked consistency for weak orders is in $\p$.
\end{corollary}

\subsection{Single-Plateaued Consistency}\label{sec:splat}

Single-plateaued preferences are a much more restrictive model than
\extsp\ %
preferences since they are essentially single-peaked except that
each preference order can %
have multiple most preferred
candidates~\cite[Chapter 5]{bla:b:polsci:committees-elections}.

Since a preference order must be strictly increasing and then %
strictly decreasing with respect to an axis %
(excluding its most preferred candidates) we can again use
the v-valley substructure. However we will need
another substructure to
prevent two candidates that are ranked indifferent in a voter's preference order 
from appearing on the same side of that voter's peak (plateau). %

\begin{definition}\label{def:nonpeak-plat}
 A preference order $v$ over a candidate set $C$ contains a
 \emph{nonpeak plateau with respect to $A$} if there exist
 candidates $a,b,c, \in C$ such that
 $a < b < c$ in $A$ and either $(a > b \sim c)$ or $(c > b \sim a)$ in $v$.
\end{definition}

We use the v-valley and nonpeak plateau substructures to state the
following lemma.

\begin{lemma}\label{lem:plat}
Let $V$ be a preference profile of weak orders. $V$ is
single-plateaued with respect to an axis $A$ if and only if no
preference order $v \in V$ contains a v-valley with respect to $A$
and no preference order $v \in V$ contains a nonpeak plateau with
respect to $A$.
\end{lemma}

\begin{proofs}
Let $C$ be a candidate set, $V$ be a preference profile of weak orders, and
$A$ be an axis.

If $V$ is single-plateaued with respect to $A$ then for every preference order
$v \in V$, $A$ can be split into segments $X$, $Y$, and $Z$ such that $v$
is strictly increasing along $X$, remaining constant along $Y$, and
strictly decreasing along $Z$. 
Since $v$ is only ever strictly decreasing along $Z$ and $Z$ is the
rightmost segment of $A$, $v$ cannot contain a v-valley with respect to $A$.
For a nonpeak plateau to exist with respect to $A$ there must exist
candidates $a,b,c \in C$ such that $a < b < c$ in $A$ and either
$(a > b \sim c)$ or $(c > b \sim a)$ in $v$.

We first consider the case of $(a > b \sim c)$ in $v$. Since $a$ is strictly
preferred to $b$ and $c$ in $v$ and both $b$ and $c$ are to the right of $a$ on
the axis
we know that both $b$ and $c$ must be in segment $Z$.
However, $v$ is strictly decreasing along $Z$, so $v$ cannot have a nonpeak
plateau of this form.

We now consider the case of $(c > b \sim a)$ in $v$. Since $c$ is strictly
preferred to
$a$ and $b$ in $v$ and both $a$ and $b$ are to the left of $c$ on
the axis %
we know that both $a$ and $b$ must be in segment $X$.
However, $v$ is strictly increasing along
$X$, so $v$ cannot have a nonpeak plateau of this form.

For the other direction we consider the case when no preference
order $v \in V$ contains a v-valley with respect to $A$
and no preference order $v \in V$
contains a nonpeak plateau with respect to $A$.

Since no preference order $v \in V$ contains a v-valley with respect to $A$, we know from
Lemma~\ref{lem:extv} that $V$ is \extsp\ with respect to $A$. 
Since we also know that no preference order $v \in V$ contains a nonpeak plateau
with respect to $A$ it is easy to see that $V$ is single-plateaued with respect
to $A$.~\end{proofs}

Since the nonpeak plateau substructure is needed in addition to the
v-valley substructure, we need to extend Construction~\ref{con:mat} so that
if a
preference order contains a nonpeak plateau with respect to an axis $A$,
then when the columns of its corresponding preference matrix are permuted according to
$A$ the matrix will contain a row with the sequence
$\cdots 1 \cdots 0 \cdots 1 \cdots$.
Notice that if a preference order ranks three %
candidates indifferent to each other below its peak (plateau)
that it will have a nonpeak plateau with respect to \emph{every} possible axis.
To handle this case in the extension to
Construction~\ref{con:mat} we need to ensure that its corresponding preference
matrix will contain a row with the sequence $\cdots 1 \cdots 0 \cdots 1 \cdots$ for
every permutation of its columns.

\begin{construction}\label{con:mat3} %
Let $V$ be a preference profile of weak orders over candidate set $C$. %
For each $v \in V$ construct a ($\|C\|-1) \times \|C\|$ matrix $X_v$.
Each column of $X_v$ corresponds to a candidate in $C$. For each candidate
$c \in C$ let $k$ be the number of candidates that are strictly
preferred to $c$ in $v$ and %
let the corresponding column
in matrix $X_v$ contain $k$ 0's starting at row one, with the 
remaining
entries filled with 1's (as in Construction~\ref{con:mat}).
The following extensions to Construction~\ref{con:mat} ensure that
if $v$ has nonpeak plateau with respect to an axis $A$ then when the columns
of $X_v$ are permuted according to $A$ it will not have consecutive ones in
rows. 
For each pair of
candidates $a,b \in C$ such that $(a \sim b)$ in
$v$, they are not the most preferred candidates in $v$, and there is no
candidate $c \in C - \{a,b\}$ such that $v$ is indifferent among
$a$, $b$, and $c$, %
then %
append three additional rows to the matrix $X_v$ %
where the column corresponding 
to $a$ is $\begin{bmatrix}0 & 1 & 1 \end{bmatrix}'$,
the column corresponding to $b$ 
 is $\begin{bmatrix}1 & 1 & 0 \end{bmatrix}'$,
each column corresponding to a candidate 
strictly preferred to $a$ and $b$ 
is $\begin{bmatrix}1 & 1 & 1 \end{bmatrix}'$,
and each column corresponding to a remaining candidate
is $\begin{bmatrix}0 & 0 & 0 \end{bmatrix}'$.
If there exist %
three %
candidates $a,b,c \in C$ such that $(a \sim b \sim c)$ %
in $v$ and they are not the most preferred candidates in $v$, then
output a matrix that has no solution.

After constructing an $X_v$ matrix for each $v \in V$, all $\|V\|$
of the
matrices are row-wise concatenated to yield a matrix $X$,
except in the case where the input resulted in a matrix with no solution.

\end{construction}

We now show how Construction~\ref{con:mat3} is applied to a preference
profile of weak orders that is single-plateaued.

\begin{example}\label{ex:sp1}\normalfont
We consider the same preference profile as in Example~\ref{ex:extsp}
and we bold the additional rows in this example. %
Let the preference order $v$ be $(a \sim c > b > e \sim d > f)$ and
the preference order $w$ be $(a > b > c > e \sim d > f)$.
\[
X_v =    \stackrel{\begin{matrix} a & b & c & d & e & f\\ \end{matrix}}{
   \begin{bmatrix}
   1 & 0 & 1 & 0 & 0 & 0\\
   1 & 0 & 1 & 0 & 0 & 0\\
   1 & 1 & 1 & 0 & 0 & 0\\
   1 & 1 & 1 & 1 & 1 & 0\\
   1 & 1 & 1 & 1 & 1 & 0\\
   \mathbf{1} & \mathbf{1} & \mathbf{1} & \mathbf{0} & \mathbf{1}&\mathbf{0}\\   
   \mathbf{1} & \mathbf{1} & \mathbf{1} & \mathbf{1} & \mathbf{1}&\mathbf{0}\\
   \mathbf{1} & \mathbf{1} & \mathbf{1} & \mathbf{1} & \mathbf{0}&\mathbf{0}\\
    \end{bmatrix}}
\hspace{0.25\textwidth}
X_w =    \stackrel{\begin{matrix} a & b & c & d & e & f\\ \end{matrix}}{
   \begin{bmatrix} 
   1 & 0 & 0 & 0 & 0 & 0\\
   1 & 1 & 0 & 0 & 0 & 0\\
   1 & 1 & 1 & 0 & 0 & 0\\
   1 & 1 & 1 & 1 & 1 & 0\\
   1 & 1 & 1 & 1 & 1 & 0\\
   \mathbf{1} & \mathbf{1} & \mathbf{1} & \mathbf{0} & \mathbf{1}&\mathbf{0}\\   
   \mathbf{1} & \mathbf{1} & \mathbf{1} & \mathbf{1} & \mathbf{1}&\mathbf{0}\\
   \mathbf{1} & \mathbf{1} & \mathbf{1} & \mathbf{1} & \mathbf{0}&\mathbf{0}\\
\end{bmatrix}}\]
We then row-wise concatenate $X_v$ and $X_w$ to construct $X$.
\[ X = 
   \stackrel{\begin{matrix} a & b & c & d & e & f\\ \end{matrix}}{
   \begin{bmatrix}
   1 & 0 & 1 & 0 & 0 & 0\\
   1 & 0 & 1 & 0 & 0 & 0\\
   1 & 1 & 1 & 0 & 0 & 0\\
   1 & 1 & 1 & 1 & 1 & 0\\
   1 & 1 & 1 & 1 & 1 & 0\\
   \mathbf{1} & \mathbf{1} & \mathbf{1} & \mathbf{0} & \mathbf{1}&\mathbf{0}\\   
   \mathbf{1} & \mathbf{1} & \mathbf{1} & \mathbf{1} & \mathbf{1}&\mathbf{0}\\
   \mathbf{1} & \mathbf{1} & \mathbf{1} & \mathbf{1} & \mathbf{0}&\mathbf{0}\\
   1 & 0 & 0 & 0 & 0 & 0\\
   1 & 1 & 0 & 0 & 0 & 0\\
   1 & 1 & 1 & 0 & 0 & 0\\
   1 & 1 & 1 & 1 & 1 & 0\\
   1 & 1 & 1 & 1 & 1 & 0\\
   \mathbf{1} & \mathbf{1} & \mathbf{1} & \mathbf{0} & \mathbf{1}&\mathbf{0}\\   
   \mathbf{1} & \mathbf{1} & \mathbf{1} & \mathbf{1} & \mathbf{1}&\mathbf{0}\\
   \mathbf{1} & \mathbf{1} & \mathbf{1} & \mathbf{1} & \mathbf{0}&\mathbf{0}\\
    \end{bmatrix}}
\hspace{0.25\textwidth}
X' = 
   \stackrel{\begin{matrix} e & b & a & c & d & f\\ \end{matrix}}{
   \begin{bmatrix}   
   0 & 0 & 1 & 1 & 0 & 0\\
   0 & 0 & 1 & 1 & 0 & 0\\
   0 & 1 & 1 & 1 & 0 & 0\\
   1 & 1 & 1 & 1 & 1 & 0\\
   1 & 1 & 1 & 1 & 1 & 0\\
   \mathbf{1} & \mathbf{1} & \mathbf{1} & \mathbf{1} & \mathbf{0}&\mathbf{0}\\   
   \mathbf{1} & \mathbf{1} & \mathbf{1} & \mathbf{1} & \mathbf{1}&\mathbf{0}\\
   \mathbf{0} & \mathbf{1} & \mathbf{1} & \mathbf{1} & \mathbf{1}&\mathbf{0}\\
   0 & 0 & 1 & 0 & 0 & 0\\
   0 & 1 & 1 & 0 & 0 & 0\\
   0 & 1 & 1 & 1 & 0 & 0\\
   1 & 1 & 1 & 1 & 1 & 0\\
   1 & 1 & 1 & 1 & 1 & 0\\
   \mathbf{1} & \mathbf{1} & \mathbf{1} & \mathbf{1} & \mathbf{0}&\mathbf{0}\\   
   \mathbf{1} & \mathbf{1} & \mathbf{1} & \mathbf{1} & \mathbf{1}&\mathbf{0}\\
   \mathbf{0} & \mathbf{1} & \mathbf{1} & \mathbf{1} & \mathbf{1}&\mathbf{0}\\
    \end{bmatrix}}\]
Next, we permute the columns of $X$ such that in each row all of the ones
are consecutive to yield $X'$.
Observe that $V$ is single-plateaued with respect to this new ordering
$e < b < a < c < d < f$ as its axis. Also notice that an axis containing $d$ and  
$e$ adjacent to each other (as seen in Example~\ref{ex:extsp})
would not correspond to an
ordering of the columns of $X$ with %
consecutive ones in rows due to the additional rows from %
the extensions
made to Construction~\ref{con:mat} in
Construction~\ref{con:mat3}.
\end{example}

Construction~\ref{con:mat} ensures that no preference order contains
a v-valley and the extensions made in Construction~\ref{con:mat3} ensure
that no preference order contains
a nonpeak plateau.
So the proof of the following theorem uses a similar argument
to the proof of Theorem~\ref{thm:main}. %
Now the presence of v-valleys \emph{or}
nonpeak plateaus, not just v-valleys, is equivalent to a row
containing the sequence $\cdots 1 \cdots 0 \cdots 1 \cdots$.

\begin{theorem}\label{thm:splat}
A preference profile $V$ of weak orders is single-plateaued consistent
if and only if
the matrix $X$,
constructed using Construction~\ref{con:mat3},
has the consecutive ones property. %
\end{theorem}

\begin{proofs}
Let $V$ be a preference profile of weak orders.
We extend the argument used by Bartholdi and
Trick~\cite{bar-tri:j:stable-matching-from-psychological-model}
and the proof of Theorem~\ref{thm:main} except in this case we
use Lemma~\ref{lem:plat} instead of Lemma~\ref{lem:extv}.

If $V$ is single-plateaued with respect to an axis $A$ then
by Lemma~\ref{lem:plat} %
we know that no
$v \in V$ contains a v-valley with respect to $A$
and no $v \in V$ contains a nonpeak plateau with respect to $A$.
When the columns of the matrix $X$ are
permuted to correspond to the axis $A$
no row will  contain the sequence
$\cdots 1 \cdots 0 \cdots 1 \cdots$
since this would correspond to a preference order that strictly decreases
and then strictly increases along the axis $A$ (a v-valley) or it would
correspond to a preference
order that has two candidates ranked indifferent appearing on the same side of its peak
(a nonpeak plateau).
Therefore $X$ has the consecutive ones property. %

If $V$ is not single-plateaued then we know from Lemma~\ref{lem:plat} that
for every possible axis
there exists a preference order $v \in V$ such that $v$ contains a
v-valley or $v$ contains a nonpeak plateau with respect to that axis.
So every permutation of the columns of $X$ will
correspond to an axis where a preference order has a v-valley or a nonpeak
plateau.
As stated in the other direction, the presence of a v-valley or a nonpeak
plateau
corresponds to a row containing the sequence
$\cdots 1 \cdots 0 \cdots 1 \cdots$.
Therefore $X$ does not have the consecutive ones property.~\end{proofs}
\begin{corollary}\label{thm:splat-consist}
Single-plateaued consistency for weak orders is in $\p$.
\end{corollary}

\subsection{Single-Peaked Consistency}\label{sec:sp}
We now present our results for the strongest domain restriction on weak orders 
that we examine,
single-peaked preferences.
Recall that a preference order is single-peaked with respect to an axis $A$ if it is
strictly increasing to a single most preferred candidate (peak) and then strictly
decreasing with respect to $A$. So we again use the v-valley substructure, but
like the previous case of single-plateaued preferences we need an additional
substructure. Even if no preference order
has a v-valley with respect to $A$ it may not be single-peaked because it is
indifferent between two candidates on the same side of its peak or
has more than one most preferred candidate.

We can handle the first condition just mentioned with the
nonpeak plateau substructure used in Section~\ref{sec:splat}, but
the second condition requires us to view \emph{any} plateau as a forbidden
substructure. %

\begin{definition}\label{def:sub-plat}%
 A preference order $v$ over a candidate set $C$ contains a
 \emph{plateau} with respect to an axis $A$ if there exist
 candidates $a,b \in C$ such that $a$ and $b$ are adjacent
 in $A$ and $(a \sim b)$ in $v$.
\end{definition}

We can now use the plateau substructure and the v-valley
substructure to state the following lemma.

\begin{lemma}\label{lem:sp}
Let $V$ be a preference profile of weak orders. $V$ is
single-peaked with respect to an axis $A$ if and only if no
preference order $v \in V$ contains a v-valley with respect to $A$ and
no preference order $v \in V$ contains a %
plateau with respect to $A$. %
\end{lemma}

\begin{proofs}
Let $C$ be a candidate set, $V$ be a preference profile of weak orders, and
$A$ be an axis.

If $V$ is single-peaked with respect to $A$ then clearly $V$ is also
single-plateaued with respect to $A$. So by Lemma~\ref{lem:plat} we know that
no preference order $v \in V$ contains a v-valley with respect to $A$ and no preference
order $v \in V$ contains a nonpeak plateau with respect to $A$. Since $V$ is
single-peaked we also know that no preference order $v \in V$ has more than one most
preferred candidate so clearly no preference order $v \in V$
contains a plateau with respect to $A$. 

For the other direction we consider the case when no preference
order $v \in V$ contains a v-valley with respect to $A$ and no
preference order $v \in V$ contains a plateau with respect to $A$.

Since no preference order $v \in V$ contains a v-valley with respect to $A$ we
know from Lemma~\ref{lem:extv} that $V$ is \extsp\ with respect to $A$.
Since we also know that  no preference order $v \in V$ contains a
plateau with respect to $A$ it is easy to see that $V$ is
single-peaked with respect to $A$.~\end{proofs}

We extend Construction~\ref{con:mat3} so that if a preference order
contains a plateau with respect to an axis $A$, then when the columns of
its preference matrix are permuted according to $A$ the matrix will contain
the sequence $\cdots 1 \cdots 0 \cdots 1 \cdots$. Since
Construction~\ref{con:mat3} already
ensures this for the case of nonpeak plateaus, our
extended construction below only needs to add a condition for plateaus that
contain the most preferred candidates in a given preference order.

\begin{construction}\label{con:mat2} %
Follow Construction~\ref{con:mat3} except add the following %
condition while constructing a preference matrix $X_v$ for each
preference order $v \in V$.

If there exist two candidates %
$a,b \in C$ such that $(a \sim b)$ in $v$
and they are the most preferred candidates in $v$,
then output a matrix that has no solution.
\end{construction}

Clearly the extension to Construction~\ref{con:mat3} above ensures that if there
are multiple most preferred candidates in a preference order then the preference
matrix constructed from that order does not have the consecutive ones
property.

When a %
preference order
has a unique most preferred candidate and
is single-plateaued, it is clearly also single-peaked.
Construction~\ref{con:mat2}
ensures that no preference order contains
more than one most preferred candidate the same way that
Construction~\ref{con:mat3} ensures that no preference order contains
three or more candidates that are all ranked indifferent to each other and
that are not the most
preferred candidates, since %
this always results in a nonpeak plateau.
So the proof of the following theorem follows from the
proof of Theorem~\ref{thm:splat}, but using Lemma~\ref{lem:sp} instead of
Lemma~\ref{lem:plat}.

\begin{theorem}\label{thm:blacksp}
A preference profile $V$ of weak orders is single-peaked consistent
if and only if
the matrix $X$,
constructed using Construction~\ref{con:mat2}
has the consecutive ones property. %
\end{theorem}

\begin{corollary}\label{thm:sp-consist}
Single-peaked consistency for weak orders is in $\p$.
\end{corollary}

\section{Conclusions and Future Work}\label{sec:conclusion}

We presented three different variants of single-peaked preferences for weak orders
and showed that each of their corresponding consistency problems %
are in \p. Since we accomplished this by using transformations to the
problem of determining if a 0-1 matrix has the consecutive ones property
we are able to apply the PQ-tree algorithm from Booth and
Lueker~\cite{boo-lue:j:pqtrees}. Using this algorithm we can
actually go further than just determining the consistency problem for
each of these variants and find \emph{all} consistent axes.
An interesting open direction is how the consecutive ones matrix problem
relates to other domain restrictions and what benefits there are
to having all consistent axes for a given preference profile.

The existential approach introduced by Lackner for single-peaked
preferences~\cite{lac:c:incomplete-sp-aaai}
has been recently applied to
other domain restrictions.
The model of single-crossing preferences~\cite{mir:j:single-crossing} was
studied in the existential model by Elkind et
al.~\cite{elk-fal-lac-obr:c:incomplete-sc} and the
model of top-monotonic preferences~\cite{bar-mor:j:top-monotonicity}
was studied in the existential model by Aziz~\cite{azi:t:top-monotonicity}.
An interesting direction for future work would %
be to apply the existential model to other domain restrictions.

Single-peaked preferences are studied because they are a simply stated and
important
domain restriction that gives insight into how the voters view the candidates
and elections with single-peaked voters have nice properties.
However, experimental study suggests that %
in real-world settings voters are often not single-peaked~\cite{mat-for-gol:c:empirical-voting-large-data},
but in this study the single-peaked results only used Black's definition
for total orders. %
It would be interesting to see if real-world datasets of weak orders
contain voters that are single-peaked, single-plateaued,
or \extsp. %
In single-peaked and nearly single-peaked elections computational problems
often become easier~\cite{fal-hem-hem-rot:j:single-peaked-preferences,fal-hem-hem:j:nearly-sp}. %
As mentioned by Lackner~\cite{lac:c:incomplete-sp-aaai} an important
open problem is to determine what computational benefits are gained when
a preference profile is \extsp\ or even nearly \extsp.
There are several
different types of nearly single-peakedness and determining if a given
preference profile is
nearly single-peaked with respect to a certain distance measure is an
interesting computational problem~\cite{erd-lac-pfa:c:nearly-sp}. It
would be interesting to see how preference profiles of weak orders impact
the complexity of nearly single-peakedness or,
as also mentioned by Lackner~\cite{lac:c:incomplete-sp-aaai}, nearly
single-peakedness in the existential model.

\section{Acknowledgments}

The author thanks Edith Hemaspaandra, Martin Lackner,
David Narv{\'a}ez, Stanis{\l}aw Radziszowski, and the anonymous
referees for their
helpful suggestions and feedback.
This work was supported in part by NSF grant no.\
CCF-1101452 and a National Science Foundation Graduate Research
Fellowship under NSF grant no.\ DGE-1102937.

\end{document}